\documentclass[rapids]{jfm}
\usepackage{amsmath,amssymb,bm}
\usepackage[usenames,dvipsnames]{color}
\usepackage{natbib}
\usepackage{graphicx}
\usepackage{gensymb}
\usepackage{bm}
\usepackage{amssymb}	
\newcommand\Rey{\mbox{\textrm{Re}}}  
\newcommand\Rew{\mbox{\textrm{Re}}_\textsc{w}}
\newcommand\Reppf{\mbox{\textrm{Re}}}
\newcommand\Rehpf{\mbox{\textrm{Re}}}
\newcommand\h{h}
\newcommand\HH{H}
\newcommand\VV{V}
\usepackage{color}

\begin{document}

\title{Turbulent-laminar patterns in shear flows without walls}

\shorttitle{Turbulent-laminar patterns in shear flows without walls}

\author[M. Chantry, L.~S. Tuckerman and D. Barkley]{Matthew Chantry$^1$, Laurette S. Tuckerman$^1$ and Dwight Barkley$^2$}
\affiliation{$^1$Laboratoire de Physique et M{\'e}canique des Milieux H{\'e}t{\'e}rog{\`e}nes (PMMH), UMR CNRS 7636; PSL - ESPCI, 10 rue Vauquelin, 75005 Paris, France; Sorbonne Universit{\'e} - UPMC, Univ. Paris 06; Sorbonne Paris Cit{\'e} - UDD, Univ. Paris 07\\
[\affilskip] $^2$Mathematics Institute, University of Warwick, CV4 7AL Coventry, United Kingdom}
\maketitle

\begin{abstract}
Turbulent-laminar intermittency, typically in the form of bands and
spots, is a ubiquitous feature of the route to turbulence in
wall-bounded shear flows. Here we study the idealised shear between
stress-free boundaries driven by a sinusoidal body force and
demonstrate quantitative agreement between turbulence in this flow and
that found in the interior of plane Couette flow -- the region excluding
the boundary layers. Exploiting the absence of boundary layers, 
we construct a model flow that uses only four Fourier modes in
the shear direction and yet robustly captures the range of
spatio-temporal phenomena observed in transition, from spot growth to
turbulent bands and uniform turbulence. The model substantially
reduces the cost of simulating intermittent turbulent structures while
maintaining the essential physics and a direct connection to the
Navier-Stokes equations.
We demonstrate the generic nature of this process by introducing stress-free
equivalent flows for plane Poiseuille and pipe flows which again
capture the turbulent-laminar structures seen in transition.
\end{abstract}


\begin{figure}
\includegraphics[width=0.92\columnwidth, clip=true]{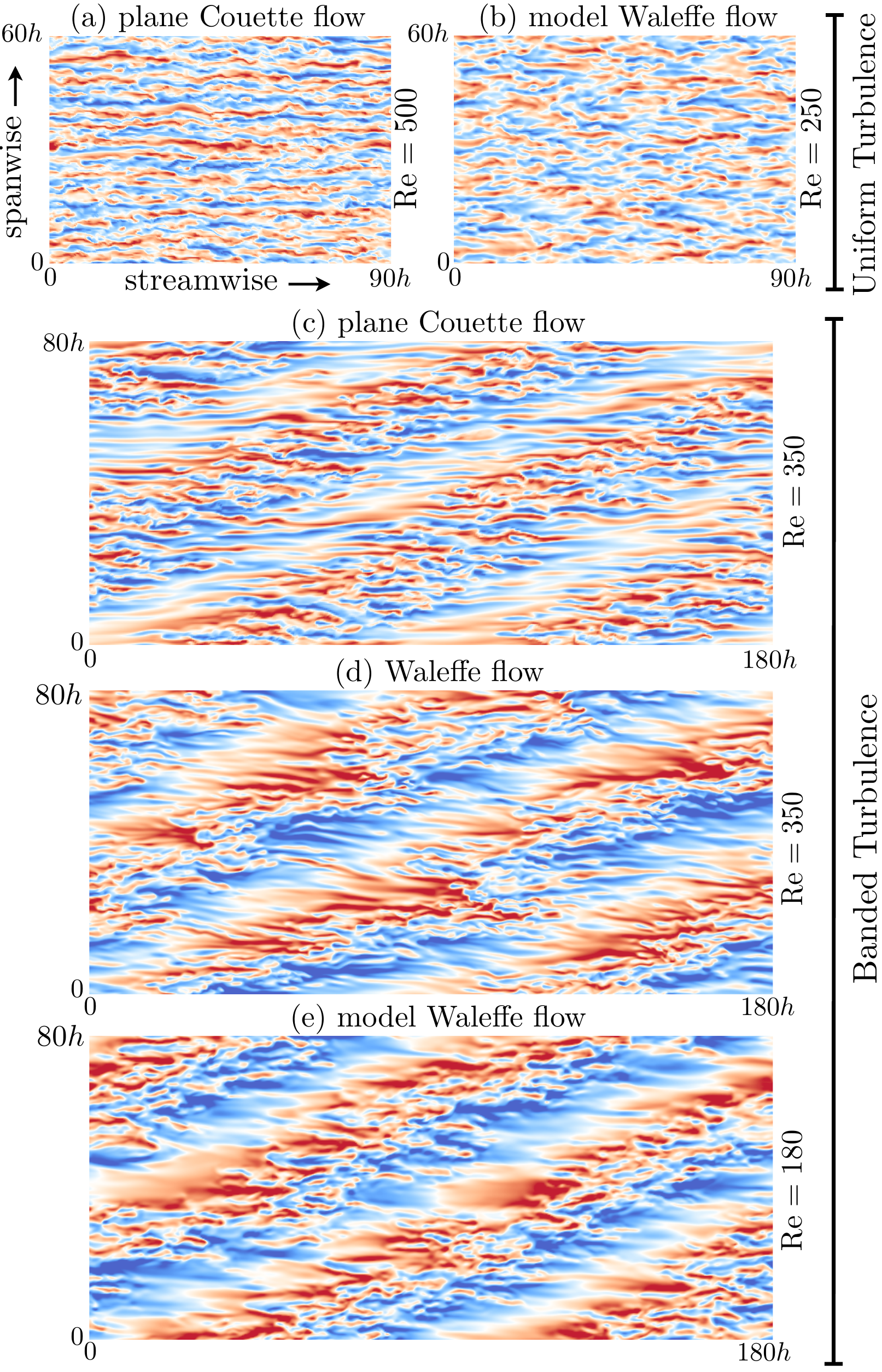}
\caption{Uniform and banded turbulence visualised by instantaneous streamwise
  velocity at the midplane, with contours from negative (blue) to positive
  (red) velocity. (a) At high $\Rey$, shear turbulence uniformly fills the
  plane Couette geometry with characteristic low- and high-speed streaks. (b)
  Comparable uniform turbulence in model Waleffe flow (introduced below).
  At lower $\Rey$, banded turbulence is observed in (c) plane Couette flow, (d) 
  Waleffe flow, and (e) model Waleffe flow.
  }
\label{fig:UT}
\end{figure}

\section{Introduction}
The onset of turbulence in wall-bounded shear flows is associated with
strong intermittency, in which turbulent and laminar flow compete on
long spatial and temporal scales. More than a mere curiosity, this
intermittency plays a key role in the route to turbulence in many
shear flows. Intermittent turbulence is well illustrated by decreasing
the Reynolds number in plane Couette flow -- the flow between parallel
rigid walls moving at different speeds. For sufficiently large
Reynolds numbers, the flow is fully turbulent and the fluid volume is
uniformly filled with characteristic streamwise streaks and rolls of
wall-bounded turbulence [Fig.~\ref{fig:UT}(a)]. With decreasing
Reynolds number, intermittency first arises as a large-scale
modulation of the turbulent streak-roll structures, eventually
resulting in persistent oblique bands of alternating turbulent and
laminar flow [Fig.~\ref{fig:UT}(c)]. As the Reynolds number is lowered
further, the percentage of turbulent flow decreases until eventually
the system returns to fully laminar flow via a percolation transition
\citep{pomeau1986front, bottin1998statistical, shi2013scale,
  manneville2015transition}. In the case of pipe flow, significant
progress has been made in understanding the various stages of the
transition process \citep{moxey2010distinct, avila2011onset,
  barkleyPipe, barkley2015}. However, in systems with two extended
directions such as plane Couette flow
\citep{prigent2002large,prigent2003long,barkley2005computational,duguet2010formation},
Taylor-Couette flow
\citep{coles1965transition,meseguer2009instability,prigent2002large}
and plane Poiseuille flow
\citep{tsukahara2014Heat,tuckerman2014turbulent}, many basic questions
remain concerning the formation and maintenance of turbulent bands and
the exact nature of the percolation transition, despite efforts to
model and understand these features
\citep{manneville2004spots,manneville2009spatiotemporal,manneville2015transition, 
  lagha2007POF,lagha2007modeling,barkley2007mean,duguet2013oblique,
  shi2013scale, SeshMann}.

Plane Couette flow (PCF)
is generally viewed as the ideal system in
which to investigate shear turbulence due to its geometric simplicity
and the constant shear rate of its laminar flow. In the turbulent
regime, however, the mean shear is far from constant. Instead it has a
low-shear core and higher-shear boundary layers associated with rigid
walls. To this end, we study a flow that surpasses PCF as an ideal
computational scenario for transition because the turbulent mean shear
is nearly constant at transitional Reynolds numbers. We show that the planar shear flow between
stress-free boundaries driven by sinusoidal body forcing reproduces
the qualitative phenomena and quantitative profiles of the core region
of PCF; it has the dual advantage of requiring far lower spatial
resolution for fully resolved simulations and lending itself to
faithful model reduction.

In fully turbulent plane Poiseuille flow (PPF), other authors have studied wall-bounded turbulence without 
walls by modelling the boundary layers combined with POD and LES frameworks \citep[e.g.][]{podvin2011synthetic,mizuno2013wall}. 
Here, we will adapt our stress-free approach to study plane Poiseuille and
pipe flows at transitional Reynolds numbers. 



\section{Waleffe flow}
\begin{figure}
{\centering \includegraphics[width=\columnwidth, clip=true]{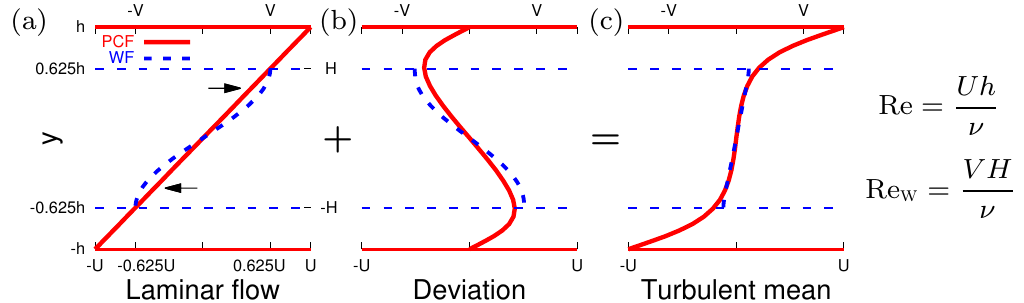} \par}
\caption{Waleffe flow seen as an approximation to the interior
of plane Couette flow.  Shown are streamwise velocity profiles for PCF
(solid/red) and WF (dashed/blue) in the uniformly turbulent regime
(PCF: $\Rey=500$ and WF: $\Rew=500$).
Plotted are (a) laminar flow, (b) deviation of mean turbulent profile from
laminar flow, and (c) mean turbulent profile.  The $y$-scale of WF is
non-dimensionalised using $\h=1.6\HH$ to align its stress-free boundaries
(dashed horizontals) with extrema of PCF deviation profile in (b). WF
velocities are likewise scaled by $U=1.6\VV$ so that both flows have the same
average laminar shear in (a).
Data are from simulations of 2000 advective time units for 
$[L_x,L_y,L_z]=[12,2,10]h$. 
}
\label{fig:mean}
\end{figure}

Plane Couette flow is generated by rigid parallel walls located at $y=\pm h$
moving with opposite velocities $\pm U$ in the streamwise direction. 
In contrast, the system we consider is 
driven by a sinusoidal body force to produce 
a laminar shear profile confined by stress-free boundary conditions
\begin{equation}u_{\rm lam}(y)= V\sin\!\left(\frac{\pi}{2} \frac{y}{\HH} \right), \qquad v (y=\pm\HH)=\left.\frac{\partial u}{\partial y}\right|_{\pm\HH}= \left.\frac{\partial w}{\partial y}\right|_{\pm\HH}=0, \end{equation}
depicted in Fig.~\ref{fig:mean}(a). 
Typically, periodic boundary conditions are imposed in the lateral streamwise, $x$, and spanwise, $z$, directions. 
The flow was first used by Tollmien to illustrate the insufficiency of an
inflection point for linear instability \citep{drazin2004hydrodynamic}.
Its simplicity derives from the stress-free boundary conditions, much as
stress-free boundaries have led to simplicity and insight in thermal
convection \citep{drazin2004hydrodynamic}.
\cite{waleffe1997self} used the flow to illustrate the self-sustaining process
and to derive a model of eight ordinary differential
equations (ODEs) capturing the essence of the process. 
Extensions of this ODE model have been derived
\citep{manneville2004spots} 
and used to measure
turbulent lifetimes \citep{moehlis,Dawes2011} as well as to find unstable
solutions \citep{moehlis2005periodic,chantryPRE,BeaumePRE}.
However, there has been little study of fully-resolved Waleffe flow itself
in the context of turbulence.
 \cite{schumacher2001evolution}
studied the lateral growth of turbulent spots and \cite{doering}
considered the bounds on energy dissipation in this system. 
Here, we 
undertake a systematic study of Waleffe flow
throughout the transitional regime.

We simulate Waleffe flow with the freely available {\tt CHANNELFLOW}
\citep{GibsonHalcrowCvitanovicJFM08,channelflow} adapted to enforce
stress-free boundary conditions.
We employ 33 Chebyshev modes in the vertical direction, $y$, and approximately 
128 Fourier modes per ten spatial horizontal units. 

We begin by comparing turbulent velocity profiles for Waleffe and plane
Couette flow, and use these to establish a scaling relationship between the
flows.
Figure~\ref{fig:mean} shows the
streamwise velocity of uniformly turbulent
flow, averaged over time and the horizontal directions, decomposed 
into the sum of the laminar profile and the deviation from
laminar.  
Lengths in WF have been scaled to align its stress-free boundaries with the
extrema of the PCF deviation profile [Fig.~\ref{fig:mean}(b)] and velocities
have been scaled to maintain the average laminar shear [Fig.~\ref{fig:mean}(a)].
WF effectively captures the interior section of PCF
-- the section between the extrema of the deviation profile,
Fig.~\ref{fig:mean}(b), or equivalently the section excluding the boundary
layers associated with no-slip walls, Fig.~\ref{fig:mean}(c).
This was first observed by \cite{waleffe2003homotopy}
for an exact solution (exact coherent structure) shared by PCF
and by another stress-free version of PCF.

The preceding paragraph implies that when treating WF as the interior of PCF, 
 WF
should be non-dimensionalised by length and velocity scales given by
$\HH=\h/1.6 =0.625\h$ and $\VV=0.625U$.
These values are not intended to be exact, since the extrema of the PCF
profiles depend on $\Rey$, although weakly over the range of interest here
(from $y/h \simeq \pm0.60$ at $\Rey=300$ to $y/h \simeq \pm0.65$ at
$\Rey=700$).
%
%
This rescaling of $y$ is almost identical to that arrived at by 
\cite{waleffe2003homotopy} through a different line of reasoning. 
A value close to this one could also be obtained from 
the extrema of low-order polynomial 
approximations, like those used for modelling by \cite{lagha2007POF,lagha2007modeling}, 
although these $y$ values would necessarily deviate from the actual values with increasing Reynolds number.
%
The effective Reynolds number for WF, comparable to that for PCF, is then

\begin{equation}\Rey \equiv \frac{Uh}{\nu} = \frac{(1.6\VV)(1.6\HH)}{\nu} = 2.56 \Rew.
\label{eq:rew}\end{equation}
where $\Rew \equiv V\HH/\nu$ is the Reynolds number usually used for WF.


Simulating Waleffe flow in large domains, we observe robust turbulent bands
emerging from uniform turbulence as the Reynolds number is decreased.
Figure~\ref{fig:UT}(d) shows such bands under conditions equivalent to those
for PCF in Fig.~\ref{fig:UT}(c). There is remarkably strong resemblance in the
broad features of the two flows.
The primary difference is that in WF, the positive (red) and negative (blue)
streaks are less distinct and are almost 
entirely
separated by the turbulent band center, while in PCF the streaks are more sharply defined
and may pass through the turbulent center. 

\begin{figure}
\includegraphics[width=\columnwidth, clip=true]{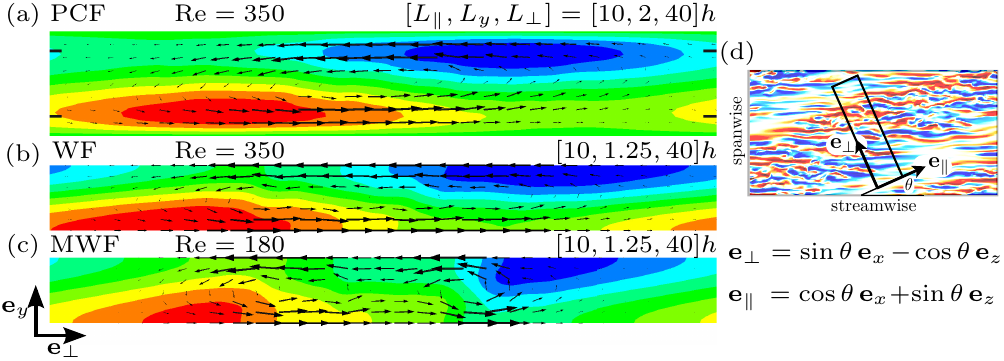}
\caption{Comparison of bands in (a) plane Couette flow, (b) Waleffe
  flow, and (c) model Waleffe flow, 
 showing the deviation from the laminar flow in a cross-sectional plane, averaged
  both in $t$ and along ${\bf{e_\parallel}}$.  The turbulent region is centered at the middle of the domain.
Through-plane flow is depicted
  by contours from negative (blue) to positive (red) and in-plane flow is
  depicted by arrows. 
  Contour levels are scaled to 10\% below extrema, PCF $\in [-0.34,0.34]$, 
  WF $\in [-0.42,0.42]$ and MWF $\in [-0.41,0.41]$.
  For
  visibility the $y$-direction in all flows has been stretched by a factor of
  3. Tic marks at $y=\pm0.625h$ in frame (a) indicate the bounds of the interior region 
  to which Waleffe flow corresponds.
 (d) Planar view of a minimal titled domain in relation to a larger
  domain.
%
 }
\label{fig:comp}
\end{figure}

For a quantitative study of the banded structure, we simulate the flows in
domains tilted by angle $\theta$ in the streamwise-spanwise plane as
illustrated in Fig.~\ref{fig:comp}(d).  
Tilted domains are the minimal flow unit to capture bands
\citep{barkley2005computational} and they provide an efficient and focused
method for quantitative analysis.
%
Domains are short ($10h-16h$) in the direction along the bands,
$\mathbf{e}_\parallel$, and long ($40h-120h$) in the direction across the
bands, $\mathbf{e}_\perp$, i.e.\ along the wavevector of the pattern.  We fix
the angle at $\theta=24\degree$, that of the bands seen in Fig.~\ref{fig:UT}.
This angle is typical of those observed in experiments and
numerical simulations of PCF in large domains
\citep{prigent2002large,duguet2010formation} and is that used in previous work
\citep{barkley2005computational,barkley2007mean,tuckerman2011patterns}
on tilted domains.


In Figs.~\ref{fig:comp}(a) and \ref{fig:comp}(b) we compare bands in Waleffe
flow to those of plane Couette flow, under equivalent conditions using the
re-scaling \eqref{eq:rew} of WF.
Mean flows are visualised in the
$(\mathbf{e}_\perp,\mathbf{e}_y$)-plane, with averages taken over the
$\mathbf{e}_\parallel$ direction and over 2000 advective time units.
The red and blue regions
indicate the flow parallel to the turbulent bands, primarily 
along the edges of the bands, 
while the arrows show circulation surrounding them.
The banded structure in Waleffe flow is almost identical to that found
in the interior of plane Couette flow. \cite{waleffe2003homotopy}
made similar observations regarding exact coherent structures
in no-slip and stress-free versions of plane Couette flow.
%
%
The main qualitative difference between the flows is the greater separation 
of the regions of positive and negative band-aligned flow in WF 
[Fig.~\ref{fig:comp}(b)]. This is a manifestation of the streak separation 
in Fig.~\ref{fig:UT}(d). 

\begin{figure}
\includegraphics[width=0.99\columnwidth, clip=true]{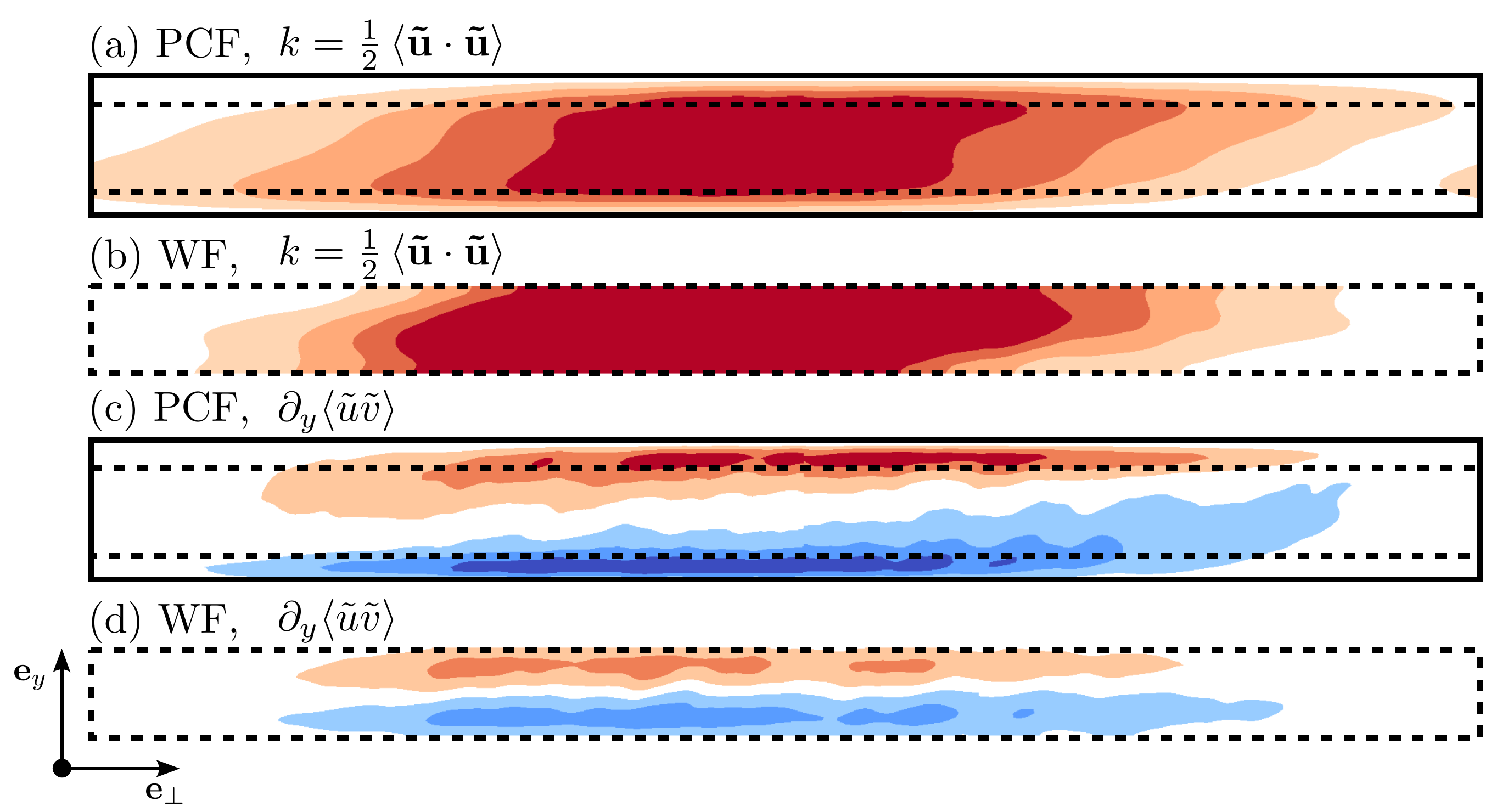}
\caption{Comparison of turbulent fluctuations, $\mathbf{\tilde{u}}$, in plane Couette flow and Waleffe flow for turbulent bands plotted in figure \ref{fig:comp}.
(a,b) Turbulent kinetic energy, $k=\tfrac{1}{2}\!\left<\mathbf{\tilde{u}}\cdot \mathbf{\tilde{u}}\right>$ for PCF, (contours $[0,0.08])$ and WF (contours $[0,0.05])$ respectively, averaged as in figure \ref{fig:comp}(a-c). (c,d) Dominant turbulent force contribution in the band-aligned direction, $\partial_y\!\left<\tilde{u}\tilde{v}\right>$ for PCF (contours $[-0.017,0.017])$ and WF (contours $[-0.017,0.017])$. Dashed lines in figures (a,c) show the bounds of the interior region to which Waleffe flow corresponds.
}
\label{fig:Forces}
\end{figure}

We also consider the fluctuations, $\mathbf{\tilde{u}}$, about the 
mean flow. In figure \ref{fig:Forces} we see that in both PCF and WF the turbulent
kinetic energy is largest in the interior. Beneath this we plot $\partial_y\!\left<\tilde{u}\tilde{v}\right>$,
which dominates the turbulent force. (See \cite{barkley2007mean}
for a full discussion of the force balance
that prevails in turbulent-laminar banded flow.)
Although
the turbulent force is very large in the near-wall regions of PCF, it mainly acts to counterbalance
the large dissipation due to the steep gradients near the walls. In the interior of PCF, both dissipative
and turbulent forces are much weaker, as is the case for the entirety of Waleffe flow.

%

%

We have surveyed the intermittency in Waleffe flow as a function of Reynolds
number.  
In the tilted domain, bands emerge from turbulence at
$\Rey \approx 640$ and turbulent patches are still observed with long lifetimes
($O(10^3)$ time units) at $\Rey \approx 250$, consistent with $\Rew = \Rey /
2.56 \approx 110$ previously found \citep[Fig.~2]{schumacher2001evolution}.
In PCF the equivalent range is 
$325 \lesssim \Rey \lesssim 420$ 
\citep{bottin1998discontinuous,bottin1998statistical,shi2013scale,tuckerman2011patterns}.

\section{Modelling Waleffe flow}
\label{sec:MWF}
Motivated by the simplicity of Waleffe flow and its ability to capture
turbulent-band formation without the boundary layers present near rigid walls,
we have developed a minimal model using only leading
Fourier wavenumbers in the shear direction $y$.
Our model of Waleffe flow (MWF) can be written as 
\begin{subequations}
\label{eqn:123}
\begin{align}
u(x,y,z)& = u_0(x,z) + &u_1(x,z)\sin(\beta y)  
+ u_2(x,z)\cos(2\beta y) + u_3(x,z)\sin(3\beta y),\\
v(x,y,z)& = &v_1(x,z)\cos(\beta y) + v_2(x,z)\sin(2\beta y) 
+ v_3(x,z)\cos(3\beta y),  \\
w(x,y,z)& = w_0(x,z) + &w_1(x,z)\sin(\beta y) 
+ w_2(x,z)\cos(2\beta y) + w_3(x,z)\sin(3\beta y), 
\end{align}
\label{eqn:123}\end{subequations}
where $\beta=\pi/2\HH$. To further simplify, we use a 
poloidal-toroidal plus mean-mode representation 
\begin{align}
\label{eqn:decomp} 
\mathbf{u} =
\nabla \times \psi(x,y,z) \mathbf{e_{y}} 
+ \nabla \times \nabla \times \phi(x,y,z) \mathbf{e_{y}} 
+  f(y) \mathbf{e}_x + g(y) \mathbf{e}_z  , 
\end{align}
where $\psi$, $f$ and $g$ match the $y$-decomposition of $u$ and $\phi$
that of $v$.  Substituting (\ref{eqn:decomp}) into the Navier-Stokes
equations and applying Fourier orthogonality in $y$, we derive our
governing equations, which are seven partial differential equations in
$(x,z,t)$ and six ODEs for the mean flows $f$ and $g$. 
The original eight-ODE model, derived by \cite{waleffe1997self} to
illustrate the self-sustaining process, is contained within the system
and can be recovered by reducing the number of modes in $y$ and
imposing a single Fourier wavenumber in $x$ and $z$.  Our model is
closely related to a series of models by Manneville and co-workers of
WF and PCF
\citep{manneville2004spots,lagha2007POF,lagha2007modeling,SeshMann}. The
first three of these attempted to capture localised dynamics with only
two modes in $y$. Turbulent bands were not spontaneously formed or
maintained; instead, spots grew to uniform turbulence.  Most recently,
and in parallel with our work, \cite{SeshMann} showed that a model of
PCF with four polynomial modes in the wall-normal direction produced
oblique bands, albeit over a narrow range of $\Rey$.

We simulate the model using a Fourier pseudo-spectral method in $(x,z)$ and
time step using backward Euler for the linear terms and Adams-Bashforth for
the nonlinear terms.  The effective low resolution in $y$ results in a
decreased resolution requirement in $(x,z)$, with only four modes needed per
spatial unit, compared with $\sim 10$ for PCF and WF.

At high $\Rey$, uniform turbulence is observed in the model
[Fig.~\ref{fig:UT}(b)], displaying the usual streamwise-aligned streaks
generated by rolls.
Streaks in MWF, as well as in WF (not shown),
typically have shorter streamwise extent than those in PCF.
Reducing $\Rey$, bands are found [Fig.~\ref{fig:UT}(e)] which are difficult to
distinguish from those in fully resolved Waleffe flow [Fig.~\ref{fig:UT}(d)];
this is also true for bands computed in the tilted domain
[Fig.~\ref{fig:comp}(b) and (c)].
The most notable qualitative difference between MWF and WF is 
the increased separation of the band-aligned flow regions and of the 
related circulating in-plane flow.
We find bands in the model for Reynolds numbers $\Rey\in[125,230]$, a large 
relative range of $\Rey$ and an
approximate rescaling of $\Rey \in[250,640]$ for fully resolved Waleffe flow.
The most likely reason for the shift in $\Rey$ is 
the lack of high-curvature modes in the wall-normal direction,
i.e. small spatial scales which would be associated with higher dissipation.
%
However, in a model for pipe flow \citep{willis2009turbulent} with few
azimuthal modes, the $\Rey$ for transition increased relative to that of fully
resolved flow. 


\begin{figure}
\includegraphics[width=0.99\columnwidth, clip=true]{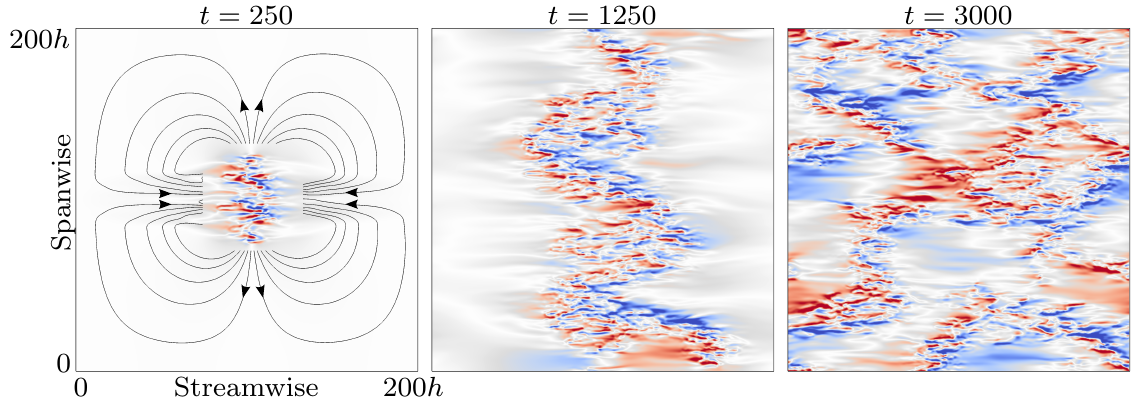}
\caption{Growth of a turbulent spot in model Waleffe flow at
  $\Rey = 160$. The flow is initialised with a poloidal vortex and subsequent
  evolution is visualised by streamwise velocity at the midplane.
  At early times ($t=250$), a large-scale quadrupolar flow dominates as shown
  by streamlines of the $y$-averaged flow (contour lines, only plotted away
  from the spot for visibility).
   By $t=1250$ bands begin to develop and form a zigzag across the
  domain. The bands continue to grow, and by $t=3000$ a complex array
  of bands fills the domain.
 }
\label{fig:spot}
\end{figure}


We investigate the formation of bands via spot growth in
the model.  As in \cite{schumacher2001evolution}, laminar flow is seeded with
a Gaussian poloidal vortex
\begin{equation} 
\mathbf{u}= \nabla \times \nabla \times 
A\exp\left[-a_x^2 x^2-a_y^2 y^2-a_z^2 z^2\right] \mathbf{e}_{y}, 
\end{equation}
here with coefficients $a_x=a_z=0.25/h$, $a_y=2/h$.
Dependence on $y$ is approximated by projecting onto the four $y$ modes
of \eqref{eqn:123}.
The developing spot in Fig.~\ref{fig:spot} matches the many facets of spot growth seen in a variety of other shear flows.
At early times
($t=250$),
growth is predominantly in the spanwise direction, as has been commonly observed \citep{schumacher2001evolution,duguet2013oblique, couliou2015large}.
An accompanying large-scale quadrupolar flow quickly develops, which we
indicate in Fig.~\ref{fig:spot}(a) by means
of streamlines of the $y$-averaged flow away from the spot. Quadrupolar flows have been reported around growing spots in PCF \citep{duguet2013oblique,couliou2015large}, in Poiseuille flow \citep{lemoult2014turbulent},and in a low-order model for PCF \citep{lagha2007POF}. 
At later times, structures develop that are recognisable as oblique bands
[compare our $t=1250$ with Fig.~1 of \cite{duguet2013oblique}]. By $t=3000$, these
structures have pervaded the whole domain.

\section{Plane Poiseuille flow}
\begin{figure}
{\centering \includegraphics[width=\columnwidth, clip=true]{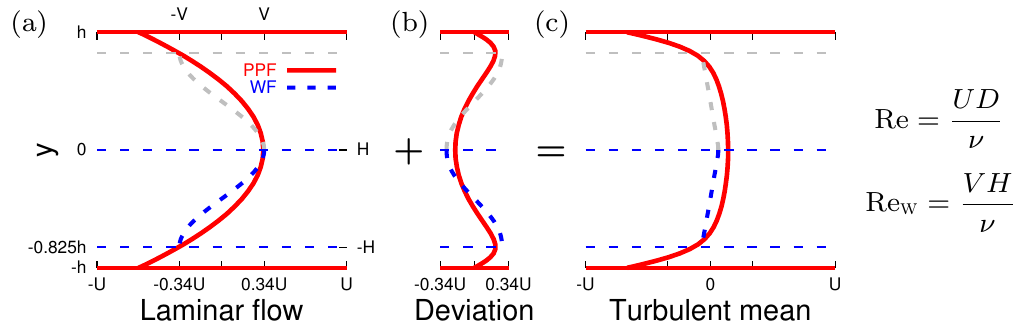} \par}
\caption{Doubled Waleffe flow seen as an approximation to the interior
  of plane Poiseuille flow.  Shown are streamwise velocity profiles for PPF
  (solid/red) and WF (dashed/blue and grey) in the uniformly turbulent regime
(PPF: $\Reppf=1800$ and WF: $\Rew=500$).
%
  Plotted are (a) laminar flow, (b) deviation of mean turbulent profile from
  laminar flow, and (c) mean turbulent profile.  The $y$-scale of WF is
  non-dimensionalised using $H=2h/0.825$ to align its stress-free boundaries
  (dashed horizontals) with extrema of PCF deviation profile in (b). WF
  velocities are likewise scaled by $V=2U/0.825^2$ so that both flows have same
  average laminar shear in (a).
  Data are from simulations of 2000 advective time units for 
  $[L_x,L_y,L_z]=[12,2,10]\h$. 
}
\label{fig:meanPPF}
\vspace*{0.5cm}
{\centering \includegraphics[width=.71\columnwidth, clip=true]{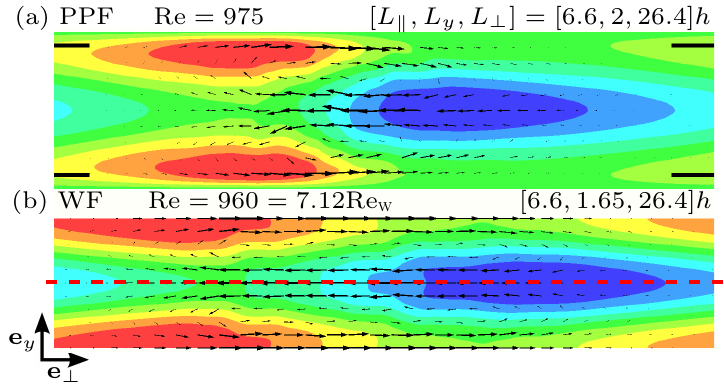} \par}
\caption{Comparison of bands between plane Poiseuille flow (top) and
  doubled Waleffe flow (bottom). Contours of streamwise velocity $[-0.4,0.4]$, and
  arrows for inplane velocity. Domain size and Reynolds number for PPF
  was chosen to match with the (rescaled) WF bands plotted in figure
  \ref{fig:comp}(b). This comparison excels near the midplane in PPF
  and confirms that PPF can be viewed as two plane Couette flows;
  compare figures \ref{fig:PPF}(a) and \ref{fig:comp}(a).  }
\label{fig:PPF}
\end{figure}

To further demonstrate the applications of this stress-free modelling
we consider plane Poiseuille flow
(PPF), generated here by enforcing constant mass flux in the horizontal directions. The laminar profile in a reference frame moving with the mean velocity is shown as the red curve of figure \ref{fig:meanPPF}(a).
A natural extension of the PCF case would be to
approximate the parabolic PPF with a cosine body forcing and stress-free boundaries.
However, such a flow develops a linear instability at $\Rey=80$, far below
expected transition. The bifurcating eigenvector is the stress-free
equivalent of the classic Tollmien-Schlichting wave of PPF, which
becomes unstable at $\Rey=5772$. To remove this unstable mode, we
enforce symmetry across the channel midplane, effectively 
juxtaposing WF (blue) with
its mirror-symmetric counterpart (grey). 
Because of this, no new simulations are necessary, since
all results concerning WF can be used, merely by using the rescaling
appropriate to PPF.  WF should now be non-dimensionalised by length
and velocity scales given by $\HH =0.825\h/2$ and $\VV=0.825^2 U /2 $.
%
The conventional PPF Reynolds number and the corresponding one for WF in this context are
\begin{equation}
\Rey\equiv \frac{U h}{\nu} = \frac{(2V)(2\HH)}{0.825^3\nu}\approx 7.12\frac{V \HH}{\nu} = 7.12 \, \Rew ,
\end{equation}
where $U$ is based on the mean Poiseuille flow. As was the case for PCF these values are not intended to be exact,
since the extrema of the PPF profiles depend on $\Rey$ (from $y/h
\simeq \pm0.78$ at $\Rey=1300$ to $y/h \simeq \pm0.86$ at
$\Rey=2400$). A ``true'' rescaling of the flow would be Reynolds
number dependent but a fixed value suffices for our purpose.
%
As in the PCF case, the length scale found by \cite{waleffe2003homotopy}
using the exact coherent structures is close to that found here
using the turbulent mean profile; 
a value within this range could also be obtained from 
the extrema of the low-order polynomials used by
\cite{lagha2007turbulent}
to model PPF.
%
Our remapped existence range for bands in Waleffe flow is $\Reppf \in
[700,1800]$ and compares well with $\Reppf \in [800,1900]$ in PPF
\citep{tuckerman2014turbulent}.  

Figure \ref{fig:PPF} shows the mean structure of turbulent bands in
PPF and in its stress-free counterpart. Excluding the boundary layers
of PPF, there is very good agreement between the structures in these
flows.
By construction, the lower half of figure \ref{fig:PPF}(b) is identical to figure
\ref{fig:comp}(b).
%
The lower half of figure \ref{fig:PPF}(a) also strongly resembles
\ref{fig:comp}(a).
The 
resemblance between turbulent bands in these two flows solidifies the
prevalent view of PPF as two PCFs \citep{waleffe2003homotopy,tuckerman2014turbulent}.

The low-order model of Waleffe flow derived for PCF in section
\ref{sec:MWF} carries over in a straightforward manner to PPF and is
therefore not shown. A five-mode model of wall-bounded PPF was derived by \cite{lagha2007turbulent} and used to study spot growth.

\section{Stress-free pipe flow}

\begin{figure}
{\centering \includegraphics[width=\columnwidth, clip=true]{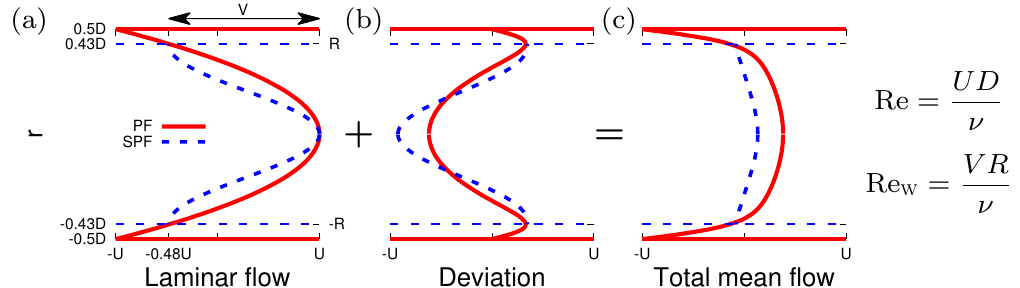} \par}
\caption{Stress-free pipe flow (SPF) seen as an approximation to the interior
  of pipe flow (PF).  Shown are streamwise velocity profiles for PF
  (solid/red) and SPF (dashed/blue) in the uniformly turbulent regime
(PF: $\Rehpf=3000$ and SPF: $\Rew=1900$).
%
  Plotted are (a) laminar flow, (b) deviation of mean turbulent profile from
  laminar flow, and (c) mean turbulent profile.  The $r$-scale of SPF is
  non-dimensionalised using $R=0.86D/2$ to align its stress-free boundaries
  (dashed horizontals) with extrema of PF deviation profile in (b). SPF
  velocities are likewise scaled by $V=2\cdot0.86^2 U$ so that both flows in (a) have the same
  average laminar shear.
  Data are from simulations of 2000 advective time units for pipes of length
  $5D$.
}
\label{fig:meanHPF}
\end{figure}

Finally, we turn to pipe
flow (PF),
the third canonical wall-bounded shear
flow, in which intermittency takes the form of puffs.
We introduce a Bessel function body force
which drives a laminar flow confined by cylindrical stress-free boundaries 
\begin{equation}
u_{z,{\rm lam}}(r)=\frac{V}{1- J_0(k'_0)} J_0 \left( k'_0 \frac{r}{R} \right), \qquad
u_r(r=R) = \left.  \frac{\partial u_z}{\partial r}  \right|_{R} =  
\left. \frac{\partial }{\partial r} \left(  \frac{u_\theta}{r}\right) \right|_{R} = 0
\end{equation}
where $k'_0\approx 3.83$ is the first non-zero root of  $J'_0$.

Simulations are conducted using {\tt openpipeflow.org} 
\citep{willis2009turbulent} 
adapted to enforce stress-free conditions on the pipe walls. As before, 
we first consider uniform
turbulence (figure \ref{fig:meanHPF}) and nondimensionalise our stress-free flow to match the turning
points in the deviation. This results in $R = 0.86\, D/2$ and $V = 0.86^2 \cdot2U$ and a Reynolds number
\begin{equation}\Rehpf =  \frac{UD}{\nu} = \frac{V\, R}{0.86^3\nu} \approx 1.57\Rew .
\end{equation}%
where $D/2$ is the pipe radius and $2U$ is the maximum laminar speed. 
Like the cosine forced version of PPF, stress-free pipe flow undergoes a linear instability at low 
Reynolds number ($\Rehpf \approx 340$), below the existence
range of turbulence. Therefore to study laminar-turbulent intermittency (here turbulent puffs) we impose the symmetry
\begin{equation} \mathbf{R}_n : \mathbf{u}(r,\theta,z) \rightarrow \mathbf{u}(r,\theta + \tfrac{2\pi}{n},z), \,\,\, n \geq 2
\end{equation}%
which stabilises the laminar flow. We will only present results from $ \mathbf{R}_3$ here but alternative 
choices (e.g. two and four) produce comparable results.

\begin{figure}
\includegraphics[width=0.99\columnwidth, clip=true]{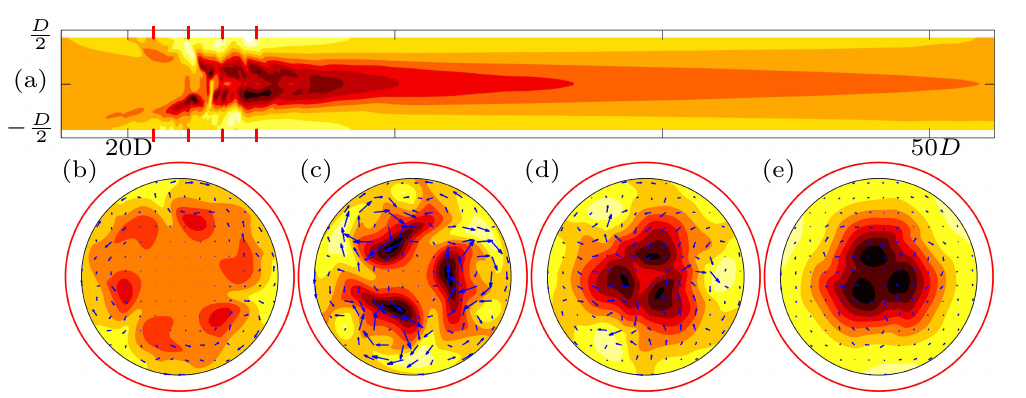}
\caption{A turbulent puff in stress-free pipe flow. (a) Streamwise velocity along the pipe (only partial pipe shown)
and (b-e) $(r,\theta)$-slices along the pipe [indicated by red lines in (a)] with arrows of in-plane velocity.
Motivated by figure \ref{fig:meanHPF} the red circles
show the location of the walls for the corresponding wall-bounded flow, 
highlighting the absence of boundary layers in the stress-free case.
Nine contours are used for streamwise velocity varying in [-0.86,0.48].}
\label{fig:Pipe}
\end{figure}

In this symmetry subspace, turbulent puffs are found for stress-free
pipe flow over a range of Reynolds numbers
$\Rehpf \in [2400,3500]$; an exemplar is plotted in figure \ref{fig:Pipe}.
For conventional rigid-wall pipe flow in this subspace, turbulent
puffs are first observed at $\Reppf\approx2400$, an increase from
$\Reppf\approx1750$ \citep{darbyshire1995transition} for no imposed
symmetry.
The structure and length scales of these puffs are comparable with their
wall-bounded counterparts. Excitation occurs upstream, [figure
  \ref{fig:Pipe}(b-c)] generating fast and slow streaks which slowly
decay downstream [\ref{fig:Pipe}(e)]. 
The success of model Waleffe
flow combined with the low-azimuthal-resolution model of
\cite{willis2009turbulent} suggests that a model with one spatial
dimension ($z$) is possible. However, the complexity of cylindrical
coordinates, particularly the coupled boundary conditions, 
prevents further work at this time.

\section{Conclusion}

Since at least the 1960s \citep[e.g.][]{coles1962interfaces} there has been interest in
understanding the ubiquitous turbulent-laminar intermittency observed
at the onset of turbulence in wall-bounded shear flows. We have
demonstrated that shear alone is the necessary ingredient for
generating these structures; the boundary layers of wall-bounded flows
are not essential. The robustness of this concept is demonstrated, not
only by turbulent bands in stress-free versions of PCF and PPF, but
also by puffs in stress-free pipe flow. Our rescaling yields
quantitative correspondence to the range of existence and the length
scales of these phenomena.  In planar geometry, we exploit the absence
of rigid walls to propose a simple four-vertical-mode model that
captures all the essential physics in the shear-dependent
direction. This provides a direct link between ODE models of the
self-sustaining process \citep{waleffe1997self} and the modelling of
turbulent-laminar coexistence. The absence of rigid walls opens the
possibility of exploring large-scale features of transitional
turbulence without the complications and numerical requirements of
sharp gradients. This should greatly facilitate the numerical study of
percolation in systems with two extended directions, while maintaining
a direct connection with the Navier-Stokes equation.


\begin{acknowledgments}
MC was supported by a grant, TRANSFLOW, provided by the Agence Nationale de la Recherche (ANR).
This work was performed using high performance computing resources provided
by the Institut du Developpement et des Ressources en Informatique
Scientifique (IDRIS) of the Centre National de la Recherche
Scientifique (CNRS), coordinated by GENCI (Grand \'Equipement National
de Calcul Intensif). 
\end{acknowledgments}

\bibliography{biblio}
\bibliographystyle{jfm}

\end{document}